\documentclass[twocolumn]{jpsj2} 
\title{Theoretical Study on Superconductivity in Boron-Doped Diamond}

\author{
Tomonori \textsc{Shirakawa}$^{1}$
\thanks{E-mail address: sirakawa@physics.s.chiba-u.ac.jp}, 
Satoshi \textsc{Horiuchi}$^{1}$, 
Yukinori \textsc{Ohta}$^{1,2}$ and 
Hidetoshi \textsc{Fukuyama}$^{3}$ 
}

\inst{$^{1}$Graduate School of Science and Technology, Chiba University, Chiba 263-8522 \\
$^{2}$Department of Physics, Chiba University, Chiba 263-8522 \\
$^{3}$Department of Applied Physics, Tokyo University of Science, Kagurazaka, Tokyo 162-8601}

\abst{We consider superconductivity in boron (B) doped 
diamond using a simplified model for the 
valence band of diamond.  We treat the effects of 
substitutional disorder of B ions by the coherent 
potential approximation (CPA) and those of the attractive 
force between holes by the ladder approximation under 
the assumption of instantaneous interaction with the 
Debye cutoff.  
We thereby calculate the quasiparticle life time, 
the evolution of the single-particle spectra due 
to doping, and the effect of disorder on the superconducting 
critical temperature $T_c$.  We in particular compare 
our results with those for supercell calculations 
to see the role of disorder, which turns out to be 
of crucial importance to $T_c$.}

\kword{boron-doped diamond, superconductivity, disorder, 
coherent potential approximation}

\begin{document}
\maketitle

\section{Introduction}

\begin{figure}[bht]
\begin{center}
\includegraphics[scale=0.36]{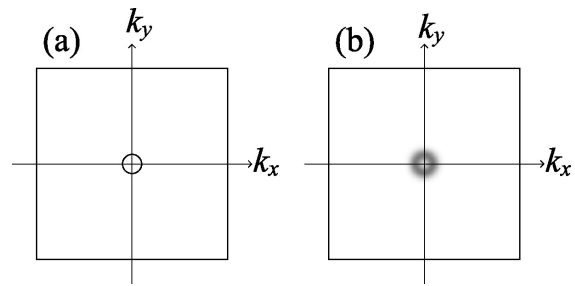}
\end{center}
\caption{Schematic representation of the Fermi surface in 
the two-dimensional analogue of the Brillouin zone for 
hole-doped diamond; 
(a) without disorder, a small Fermi surface is present, 
but (b) with disorder, the Fermi surface is smeared out 
by scattering.}
\label{fs}
\end{figure}

Recent discovery of superconductivity in boron (B) 
doped diamond \cite{ekimov,takano1} has renewed 
interest in physics of doped semiconductors. 
Inelastic x-ray scattering experiment \cite{hoesch} 
indicates that the doped carriers are strongly 
coupled with bond-stretching phonons of 
$\omega\simeq 164$ meV associated with the covalent 
bonds into which the holes are doped.  
The relatively high critical temperature of 
$T_c\simeq 10$K \cite{takano2,umezawa} is 
remarkable in view of the expected small density 
of states at the Fermi energy $\varepsilon_{\rm F}$ 
due to small carrier density \cite{mukuda} as well 
as the existence of strong disorder intrinsic 
to the randomness in doped semiconductors 
\cite{baskaran1,baskaran2}.  
In particular, the latter is deduced from the 
electrical resistivity measurement 
\cite{bustarret,sidorov,ishizaka1} 
and the line shape of quasiparticles observed by 
the angle-resolved photoemission spectroscopy (ARPES) 
\cite{yokoya}, indicating that the life time of 
electrons $\tau$ is very short with the characteristic 
parameter of $\varepsilon_{\rm F}\tau\lesssim 1$ 
(throughout this paper we set $\hbar=k_{\rm B}=1$) and 
violates the Ioffe-Regel criterion for coherent 
Bloch-like transport \cite{baskaran1,baskaran2,ishizaka1}.  
This small value of $\varepsilon_{\rm F}\tau$ implies 
that the Fermi surface is not well-defined in 
momentum space but is blurred by scattering 
(see Fig.~\ref{fs}); the superconductivity in B-doped 
diamond may therefore be referred to as 
``superconductivity without Fermi surface'' 
\cite{fukuyama0}.  

The strong scattering potential of the impurity B ions may 
be responsible not only for such disorder-induced anomalous 
features but also for the basic electronic states.  
The energy level of a single B impurity in diamond is very 
deep, locating at 0.37 eV above the top of the valence 
band of diamond \cite{ramdas,kim}, and, in the presence of 
such a strong scattering potential, the simple picture of 
the `rigid-band shift' of the Fermi level due to doping (or 
the virtual crystal approximation) based on the 
first-principles LDA (local-density approximation) band 
calculations \cite{boeri,lee,blase} with the adjusted 
chemical potential may not necessarily be justified.  
In fact, some spectroscopic experimental data 
\cite{nakamura1,nakamura2,wu} seem to suggest the presence 
of non-rigid-band--like features reminiscent of the 
impurity states even in the metallic regime of B-doped 
diamond, which has been focused upon by Baskaran 
\cite{baskaran1,baskaran2}.  Hence, evolution of the 
impurity states due to doping should be accessed with 
more care.  

Another key feature is the observed very small carrier 
concentration.  A recent analysis of the NMR spectra 
\cite{mukuda} suggests that the real doping rate of 
carriers (or the number of carriers per site) is at 
most only $\sim 1$\%, which differs very much from the 
nominal doping rates of $\sim 5\%$, and that this is 
due to the formation of boron-hydrogen (BH) complex; 
i.e., H atoms inevitably introduced in the synthetic 
processes of the materials absorb the holes introduced 
by B impurities.  
This possibility has been suggested recently both 
theoretically and experimentally;  an LDA calculation 
has shown that the B ions occupy the carbon sites 
substitutionally and the H ions sit on the interstitial 
positions between B ions and the neighboring carbon 
ions, \cite{oguchi} and based on the assumption of the 
presence of these two types of B ions, the observed 
NMR spectra can be analyzed consistently \cite{mukuda}.  
We should therefore reexamine the electronic states 
of B-doped diamond, in particular, in the regime of 
low carrier concentration of $\sim 1\%$, starting 
from the dilute limit of B doping.  

In this paper, we will first study the electronic states 
of this highly disordered system, based on a simplified 
model for the valence band of diamond by use of the 
coherent potential approximation (CPA) for treating 
the effects of the substitutional disorder of B ions.  
We thereby calculate the density of states, life time, 
and evolution of the single-particle spectra due to 
doping, the results of which are compared with available 
experimental data including ARPES spectra.  

We will also calculate the superconducting pairing 
susceptibility based on the ladder approximation under 
the assumption of instantaneous interaction and 
estimate the doping dependence of the critical 
temperature $T_c$.  We thereby want to clarify the 
effects of disorder on the superconductivity in such 
unusual situations as described above.  
We will in particular compare our results with those 
for supercell calculations (i.e., the periodic 
arrangement of B ions) to see the role of disorder, 
which turns out to be of crucial importance.  
Here we do not pursue processes of the attractive 
interaction in detail, though we assume it is 
due to the electron-phonon interaction associated 
with bond-stretching modes with very high frequency, 
as has been indicated both experimentally and 
theoretically \cite{hoesch,boeri,lee,blase,xiang}.  
Preliminary results of our work have been presented 
in ref.~\cite{ohta}

This paper is organized as follows.  In \S 2, we 
present our model and method of calculations.  
In particular, we derive the pairing susceptibility 
within the framework of CPA.  
In \S 3, we present our results of calculations 
for the density of states, life time, single-particle 
spectra, and superconducting critical temperature, 
which are compared with available experimental data 
in \S 4.  Discussions are given in \S 5, and we 
summarize our work in \S 6.  

\section{Method of Calculations}

\subsection{Model}

In order to extract the effects of disorder on the 
electronic states near the top of the valence bands 
of diamond, we assume, for the sake of simplicity, 
the tight-binding model based on the simple cubic 
structure.  By this simplification, the degeneracy 
of three bands at the top of the valence bands of 
diamond is ignored.  However, we believe that the 
essential features of the effects of disorder can 
be taken into account in this simple model.  

Adding the random potential $\Delta$ at B sites 
and the on-site attractive interaction $-V$ $(V>0)$, 
our model may be written by the Hamiltonian 
\begin{equation}
\mathcal{H}=\sum_{{\mib k}\sigma}\,
(\varepsilon_{\mib k}-\mu)\,
c^\dagger_{{\mib k}\sigma}c_{{\mib k}\sigma}
+\Delta\sum_{i\in{\rm B}}\sum_\sigma n_{i\sigma}
-V\sum_in_{i\uparrow}n_{i\downarrow}
\label{hamiltonian}
\end{equation}
with 
\begin{equation}
\varepsilon_{\mib k}=t\,(\cos k_xa+\cos k_ya+\cos k_za),
\label{band}
\end{equation}  
where $n_{i\sigma}$ is the number operator at 
site $i$ and spin $\sigma$ and 
$c^\dagger_{{\mib k}\sigma}$ creates an electron 
with momentum ${\mib k}$ and spin $\sigma$, 
$\mu$ is the chemical potential.  

\begin{figure}[htb]
\begin{center}
\includegraphics[scale=0.32]{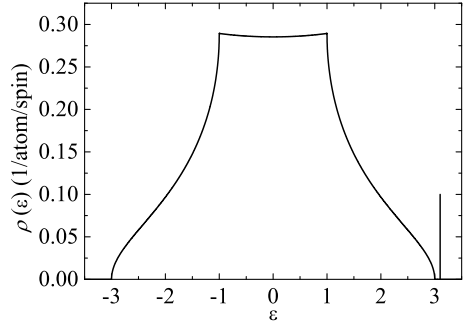}
\end{center}
\caption{Model density of states for the tight-binding 
band on the three-dimensional simple cubic lattice 
without disorder.  
Energy level of a single impurity is also shown.}
\label{modeldos}
\end{figure}

Hereafter, we assume $t=1$ as the unit of energy 
unless otherwise indicated; to reproduce the width 
of the valence band of diamond $22$ eV, this 
choice corresponds to taking $3.7$ eV as 
the unit of energy.  The effective mass at the 
top of the valence band of diamond is roughly 
reproduced by this choice.  

We determine the value of $\Delta$ so as to reproduce 
the energy level of B ions in the dilute limit, which 
is experimentally known to be located at $0.37$ eV 
above the top of the valence band of pure diamond 
\cite{ramdas,kim}; 
we obtain the value $\Delta=2.30$, which is used 
throughout the present work.  
The density of states of this model is shown 
in Fig.~\ref{modeldos} with the energy level of 
a single B impurity.  
The sites with $\Delta$ are distributed randomly over 
the lattice sites of the simple cubic structure; 
this substitutional disorder is treated by CPA.  
We define the concentration of B ions as 
$c_{\rm B}=N_{\rm B}/N$ where $N_{\rm B}$ is the 
number of B sites and $N$ is the total number of 
sites in the system.  

\subsection{Coherent Potential Approximation}

\begin{figure}[hbt]
\begin{center}
\includegraphics[scale=0.55]{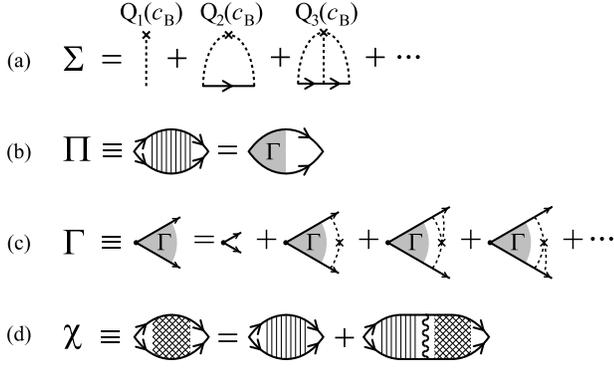}
\end{center}
\caption{(a) The self-energy $\Sigma$ by CPA.  Dotted 
and solid lines represent the perturbation potential 
$\Delta$ and Green's function, respectively, and 
$\times$ represents $Q_S(c_{\rm B})$.  
(b) Diagram for the pairing susceptibility $\Pi$ 
given in eq.~(\ref{pi2}).  
(c) Diagram for determining the vertex function 
$\Gamma$ in eq.~(\ref{gamma1}).  
(d) Diagram for the pairing susceptibility $\chi$ 
by the ladder approximation eq.~(\ref{super}).  
Wavy line represents the attractive interaction $V$.}
\label{diagram}
\end{figure}

Let us assume the absence of the attractive 
interaction, i.e., $V=0$, for the moment.  
In CPA, the Green's function is defined with the 
momentum-independent self-energy function 
$\Sigma(i\varepsilon_n)$ as 
\begin{equation}
G^{-1}({\mib k},i\varepsilon_n)=i\varepsilon_n
-\varepsilon_{\mib k}+\mu-\Sigma(i\varepsilon_n)
\label{greenfunc}
\end{equation}
where $\varepsilon_{\mib k}$ is the band structure 
without impurities eq.~(\ref{band}), $\mu$ is 
chemical potential, and $\varepsilon_n=(2n+1)\pi/\beta$, 
with $\beta=1/T$.  
The self-energy function in the presence of the 
substitutional disorder is given as the polynomial 
series expansion \cite{VKE,yonezawa,elliott} as 
shown in Fig.~\ref{diagram}(a): 
\begin{subequations}
\begin{align}
\Sigma(i\varepsilon_n)&=
\Delta\sum_{S=1}^\infty Q_S(c_{\rm B})\,
\Delta^{S-1}\,\Theta^{S-1}(i\varepsilon_n) \\
&=\frac{c_{\rm B}\Delta}{1-\Theta(i\varepsilon_n)
(\Delta-\Sigma(i\varepsilon_n))}
\end{align}
\label{seriesexp}
\end{subequations}
where 
\begin{equation}
\Theta(i\varepsilon_n)=\frac{1}{N}
\sum_{\mib k}G({\mib k},i\varepsilon_n)
\label{theta}
\end{equation}
and $Q_S(c_{\rm B})$ is the polynomial of $c_{\rm B}$.  
In the following, we use 
abbreviations $\Sigma=\Sigma(i\varepsilon_n)$ 
and $\Theta=\Theta(i\varepsilon_n)$.  

The CPA equation for $\Sigma$ 
\begin{equation}
\Theta\Sigma^2+(1-\Delta\Theta)\Sigma-c_{\rm B}\Delta=0.
\label{cpaeq}
\end{equation}
is solved numerically by use of eq.~(\ref{band}).  

\subsection{Pairing Susceptibility by CPA} 

We consider the pairing (or Cooper-pair) 
susceptibility in the absence of $V$ defined as 
\begin{equation}
\Pi({\mib q},i\omega_m)= 
\int_0^\beta{\rm d}\tau\,e^{i\omega_m\tau} 
\left<T_{\tau}b_{\mib q}^{\dagger}(\tau)
b_{\mib q}\right>
\label{pi1}
\end{equation}
with $b_{\mib q}=\sum_{\mib k}
c_{{-\mib k}-{\mib q}/2,\downarrow}
c_{{\mib k}-{\mib q}/2,\uparrow}$ 
and $\omega_m=2\pi m/\beta$.  
Following the CPA theory for the magnetic 
susceptibility \cite{fukuyama1,fukuyama2}, 
we find that eq.~(\ref{pi1}) is given by 
\begin{equation}
\Pi({\mib q},i\omega_m)=T\sum_{n}
\sum_{\mib k}G\,G_+\,\Gamma 
\label{pi2}
\end{equation}
with 
\begin{subequations}
\begin{align}
&G=G({\mib k},i\varepsilon_n)\\
&G_+=G(-{\mib k}+{\mib q},-i\varepsilon_n+i\omega_m)\\
&\Gamma=\Gamma({\mib q},i\varepsilon_n,i\omega_m),
\end{align}
\end{subequations}
which is illustrated in Fig.~\ref{diagram}(b).  
The vertex function $\Gamma$ is determined by 
the equation (see Fig.~\ref{diagram}(c)): 
\begin{equation}
\begin{split}
\Gamma&=1+\Gamma A\\
&\times\sum_{S=2}^\infty\Delta^SQ_S 
\left(\Theta^{S-2}+\Theta^{S-3}
\Theta_++\cdots+\Theta^{S-2}_+\right) 
\end{split}
\label{gamma1}
\end{equation}
with 
\begin{subequations}
\begin{align}
&\Theta_+=\Theta(-i\varepsilon_n+i\omega_m)\\
&A=A({\mib q},i\varepsilon_n,i\omega_m)
=\frac{1}{N}\sum_{\mib k}GG_+ .
\end{align}
\label{A}
\end{subequations} 
Equation (\ref{gamma1}) is solved for $\Gamma$ 
as 
\begin{equation}
\Gamma=\left(1-A\frac{\Sigma-\Sigma_+}
{\Theta-\Theta_+}\right)^{-1} 
\label{gamma2}
\end{equation}
with 
\begin{equation}
\Sigma_+=\Sigma(-i\varepsilon_n+i\omega_m).  
\end{equation}
We thus obtain 
\begin{equation}
\Pi({\mib q},i\omega_m)=T\sum_n
A({\mib q},i\varepsilon_n,i\omega_m)\,
\Gamma({\mib q},i\varepsilon_n,i\omega_m)
\label{pi3}
\end{equation}
for the pairing susceptibility in the absence of 
attractive interactions.  
Since we have 
\begin{subequations}
\begin{align}
&A(0,i\varepsilon_n,0)=
\frac{\Theta(-i\varepsilon_n)-\Theta(i\varepsilon_n)}
{2i\varepsilon_n+\Sigma(-i\varepsilon_n)-\Sigma(i\varepsilon_n)}\\
&\Gamma(0,i\varepsilon_n,0)=
\frac{2i\varepsilon_n+\Sigma(-i\varepsilon_n)-\Sigma(i\varepsilon_n)}
{2i\varepsilon_n}
\end{align}
\end{subequations} 
from eqs.~(\ref{A}) and (\ref{gamma2}), we see from 
eq.~(\ref{pi3}) that the pairing susceptibility for 
$\mib{q}=0$ and $i\omega_m=0$ is 
given as follows; 
\begin{subequations}
\begin{align}
\Pi(0,0)
&=T\sum_n
\frac{\Theta(-i\varepsilon_n)-\Theta(i\varepsilon_n)}
{2i\varepsilon_n}\\
&={1\over 2}\int\!{\rm d}\varepsilon\,
\rho(\varepsilon)\,{1\over{\varepsilon-\mu}}
\tanh\Big({{\beta(\varepsilon-\mu)}
\over{2}}\Big)
\end{align}
\label{bcspi}
\end{subequations}
where 
\begin{equation}
\rho(\varepsilon)=-\frac{1}{\pi}\lim_{\delta\to 0^+}{\rm Im}\,
\frac{1}{N}\sum_{\mib k}G({\mib k},\varepsilon-\mu+i\delta) .
\end{equation}
We thus find that the effects of disorder in the 
present approximation appear only through the density 
of states in disordered systems, $\rho(\varepsilon)$.  

We should note that the vertex correction $\Gamma$ 
is essential here; if we take into account the effects 
of disorder only from the self-energy via $A$, the 
value of the pairing susceptibility $\Pi(0,0)$ becomes 
very small (reflecting the effects of disorder too 
strongly), but the vertex correction, which becomes 
very large around the Fermi energy, recovers the value 
of $\Pi(0,0)$, to result in the BCS form of the pairing 
susceptibility eq.~(\ref{bcspi}).  
Thus, the effects of disorder, even though it is strong, 
can be renormalized solely into the density of states 
$\rho(\varepsilon)$.  This cancellation is just as 
in the case of the calculation of the magnetic 
susceptibility \cite{fukuyama1,fukuyama2}.  

In the presence of the attractive interaction, $V>0$, 
the pairing susceptibility at $\mib{q}=0$ and $i\omega_m=0$, 
$\chi(T)$, is given in the ladder approximation by 
the process shown in Fig.~\ref{diagram}(d), which 
results in  
\begin{equation}
\chi(T)=\frac{\Pi(0,0)}{1-V\Pi(0,0)}.  
\label{super}
\end{equation}
The critical temperature $T_c$ is determined from 
the condition $V\Pi(0,0)=1$ at which $\chi(T)$ 
diverges: i.e., 
\begin{equation}
\frac{1}{V}={1\over 2}\int\!{\rm d}\varepsilon\,
\rho(\varepsilon)\,{1\over{\varepsilon-\mu}}
\tanh\Big({{\beta(\varepsilon-\mu)}\over{2}}\Big).  
\label{supertc}
\end{equation}
Note that our result for $\Pi(0,0)$ can be considered 
as an extension of Anderson's theorem \cite{anderson} 
which states that $T_c$ of an isotropic $s$-wave 
superconductor is not affected by nonmagnetic impurities 
if the density of states is unchanged by scattering.  
The extension is that, within CPA, even the strong 
disorder that deforms the density of states does not 
affect the validity of the BCS expression 
eq.~(\ref{bcspi}) for the pairing susceptibility.  

\section{Results of Calculations}

\subsection{Density of States and Self-energy}

\begin{figure}[htb]
\begin{center}
\includegraphics[scale=0.47]{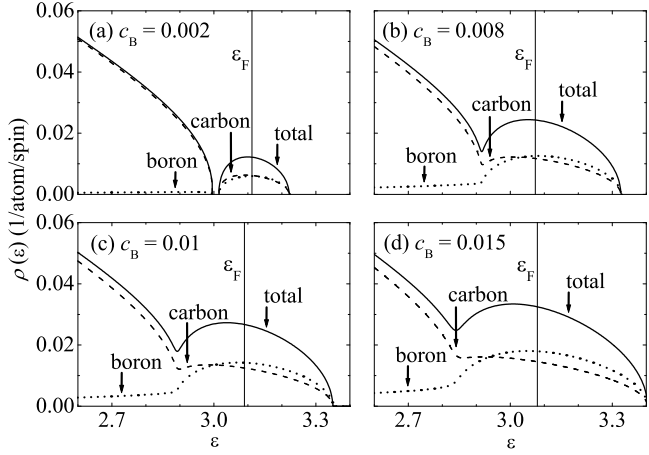}
\end{center}
\caption{Density of states $\rho(\varepsilon)$ 
near the Fermi energy calculated by CPA.  
Local densities of states for carbon and 
boron sites are also shown.  
The vertical line indicates the Fermi energy 
$\varepsilon_{\rm F}$ in the absence of 
compensation.}
\label{figdos}
\end{figure}

Calculated results for the density of states are shown 
in Fig.~\ref{figdos}.  
We find that the impurity states broaden and get 
weight continuously as the doping rate is increased.  
The broadening is due to the large imaginary part of 
the self-energy around the Fermi energy (see below).  
The impurity states are separated from the top of the 
valence band up to the doping rate of 
$c_{\rm B}\simeq 0.2$\% (see Fig.~\ref{figdos}(a)), 
but are merged into the top of the valence band 
with increasing doping.  The weight of the impurity 
states corresponds to the doping concentration.  
We assume the absence of compensation hereafter; 
i.e., the doping rate of holes is equal to $c_{\rm B}$.  
Then, the Fermi energy $\varepsilon_{\rm F}$ 
at $T=0$ K (then $\mu=\varepsilon_{\rm F}$) is 
determined by 
\begin{equation}
\int_{-\infty}^{\varepsilon_{\rm F}}\!
\rho(\varepsilon)\,{\rm d}\varepsilon
=1-\frac{c_{\rm B}}{2}.
\end{equation}
Thus, the impurity states, if they are detached from 
the valence bands, are at half filling, as pointed 
out in ref.~\cite{baskaran1}.  

\begin{figure}[htb]
\begin{center}
\includegraphics[scale=0.38]{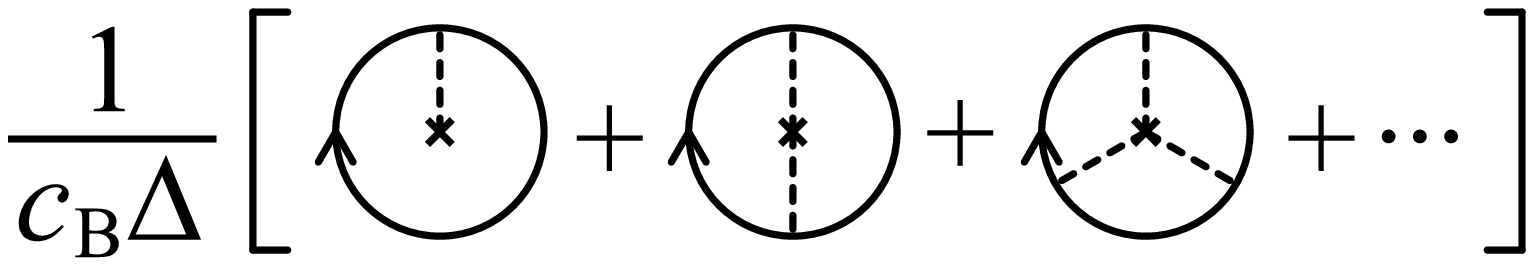}
\end{center}
\caption{(a) Diagram for defining the local density 
of states at B site, $\rho_{\rm B}(\varepsilon)$.  
Dotted and solid lines represent the perturbation 
potential $\Delta$ and Green's function, respectively, 
and $\times$ represents $Q_S(c_{\rm B})$.}
\label{localdos}
\end{figure}

The local density of states for B sites may be 
defined as in Fig.~\ref{localdos} and is given by 
\begin{subequations}
\begin{align}
\rho_{\rm B}(\varepsilon)&=-\frac{c_{\rm B}}{\pi}{\rm Im}
\frac{1}{c_{\rm B}\Delta}\sum_{S=1}^\infty
Q_S(c_{\rm B})\Delta^S\Theta^S(\varepsilon-\mu+i\delta)
\\
&=-\frac{c_{\rm B}}{\pi}
{\rm Im}\frac{\Theta(\varepsilon-\mu+i\delta)}
{1-(\Delta-\Sigma(\varepsilon-\mu+i\delta))
\Theta(\varepsilon-\mu+i\delta)}
\end{align}
\end{subequations}
with $\delta\to 0^+$.  The local density of states 
for carbon sites may be defined as 
$\rho_{\rm C}(\varepsilon)
=\rho(\varepsilon)-\rho_{\rm B}(\varepsilon)$.  
The calculated results are shown also in Fig.~\ref{figdos}, 
where we find that the boron component is 
concentrated on the impurity states but the carbon 
component in the states is as large as the boron 
component.  

\begin{figure}[htb]
\begin{center}
\includegraphics[scale=0.47]{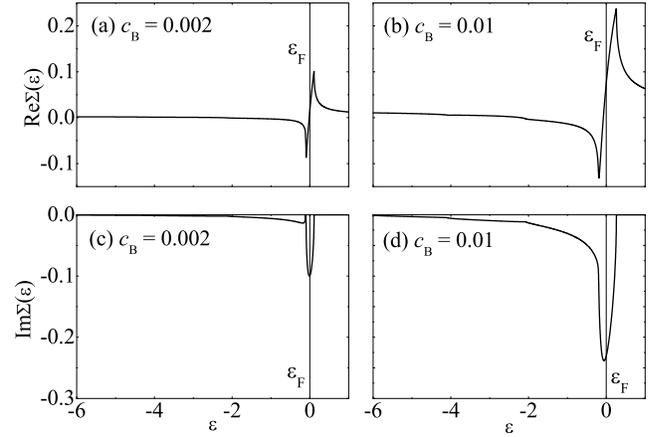}
\end{center}
\caption{Calculated real (upper panels) and 
imaginary (lower panels) parts of the self-energy 
$\Sigma(\varepsilon)$ by CPA.  
The vertical line indicates the Fermi energy in 
the absence of compensation.}
\label{figselfenergy}
\end{figure}

The real and imaginary parts of the self-energy 
$\Sigma(\varepsilon)$ are shown in Fig.~\ref{figselfenergy}, 
where we find that there is a large imaginary part 
$-{\rm Im}\Sigma(\varepsilon)$ in the impurity states, 
which are responsible for the broadening of the density 
of states of the impurity states, as well as for the 
short life time of electrons (see below).  

\subsection{Life Time and Mean-Free-Path}

\begin{figure}[thb]
\begin{center}
\includegraphics[scale=0.80]{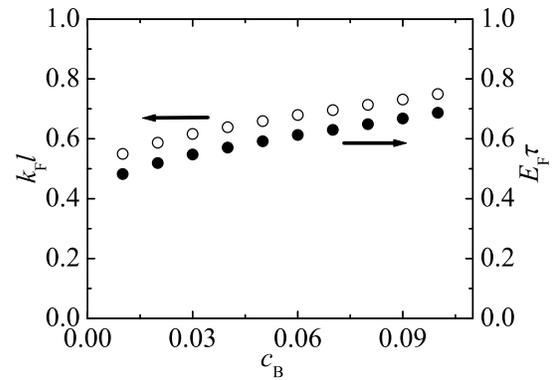}
\end{center}
\caption{Calculated life time $E_{\rm F}\tau$ 
($\bullet$) and mean-free-path $k_{\rm F}\ell$ ($\circ$) 
as a function of $c_{\rm B}$.}
\label{lifetime}
\end{figure}

The life time $\tau$ at the Fermi energy 
$\varepsilon_{\rm F}$ may be defined as 
\begin{equation}
\tau=-{1\over{2\,{\rm Im}\,
\Sigma(\varepsilon_{\rm F}+i\delta)}}
\end{equation}
with $\delta\rightarrow 0^+$, the value of which 
is determined directly from the imaginary part of 
the self-energy calculated at $\varepsilon_{\rm F}$ 
in the presence of disorder.  By use of this $\tau$, 
the mean-free-path $\ell=v_{\rm F}\tau$, $E_{\rm F}\tau$, 
and $k_{\rm F}\ell$ may be defined with 
$v_{\rm F}$, $k_{\rm F}$, and $E_{\rm F}$ being 
the Fermi velocity, Fermi wavevector, and Fermi 
energy, respectively, corresponding to the same 
carrier number but in the absence of disorder.  

The calculated results for $E_{\rm F}\tau$ 
and $k_{\rm F}\ell$ are shown in Fig.~\ref{lifetime} 
as a function of $c_{\rm B}$, where we find that 
$\tau$ and $\ell$ are indeed very short, irrespective 
of the doping rates, which comes from the large 
imaginary part of the self-energy at the Fermi energy 
(see Fig.~\ref{figselfenergy}).  
At $c_{\rm B}=0.01$, e.g., we have the values 
$E_{\rm F}\tau\simeq 0.48$ and 
$k_{\rm F}\ell\simeq 0.55$.  

\begin{figure}[tbh]
\begin{center}
\includegraphics[scale=0.95]{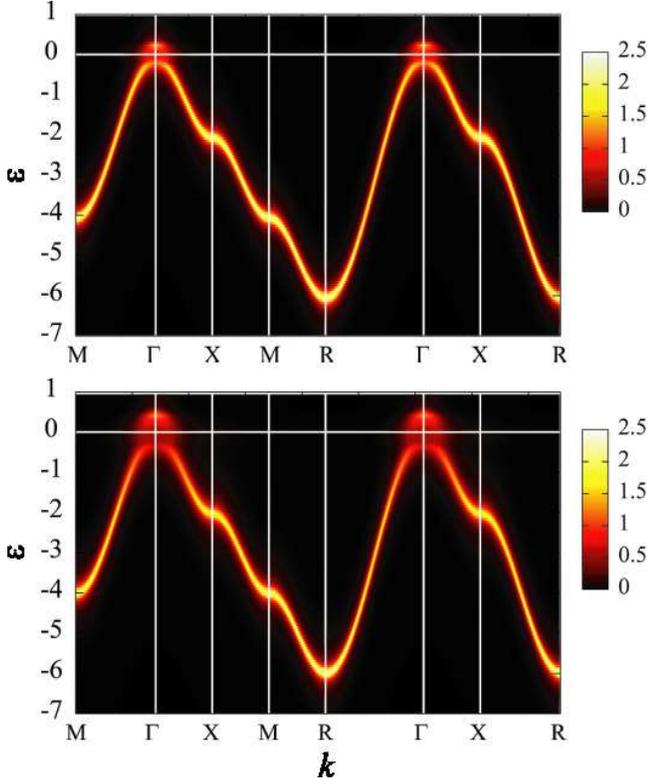}
\end{center}
\caption{(Color online) Calculated results for the 
single-particle spectra $\rho({\mib k},\varepsilon)$ 
at $c_{\rm B}=0.01$ (upper panel) and 
$c_{\rm B}=0.03$ (lower panel).  
The artificial broadening of $\delta=0.13$ is 
included in the spectra for the illustrative purpose.} 
\label{figdispersion}
\end{figure}

\subsection{Single-Particle Spectra}

The calculated single-particle spectra are shown 
in Fig.~\ref{figdispersion}, where we illustrate 
the quantity 
\begin{equation}
\rho({\mib k},\varepsilon)=-{1\over\pi}
\,{\rm Im}\,G({\mib k},\varepsilon+i\delta), 
\end{equation}
by which we can envisage the broadening of 
the band structure due to disorder in the 
Brillouin zone as well as the location of 
spectral weight due to the impurity states.  
We find that the impurity states are clearly 
seen just above the top of the valence band.  
We note that the impurity states are fairly 
localized in momentum space around the $\Gamma$ 
point even at very low doping rates.  
This result will be compared later with the 
results of the ARPES experiment \cite{yokoya}.  

\subsection{Superconducting Critical Temperature}

\begin{figure}[tbh]
\begin{center}
\includegraphics[scale=0.38]{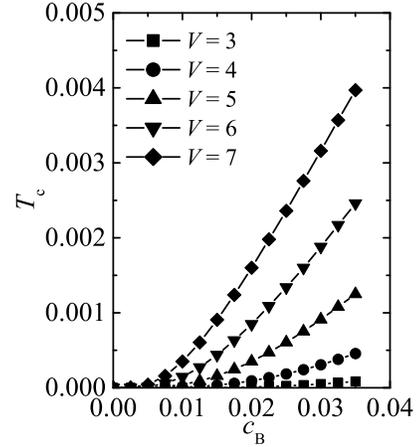}
\end{center}
\caption{Superconducting critical temperature $T_c$ 
as a function of $c_{\rm B}$ calculated by CPA.}
\label{cpatc1}
\end{figure}

The superconducting critical temperature $T_c$ is 
calculated from eq.~(\ref{supertc}) together with 
the chemical potential $\mu$ determined from 
\begin{equation}
\int_{-\infty}^\infty\!{\rm d}\varepsilon\,
f(\varepsilon)\rho(\varepsilon)=1-\frac{c_{\rm B}}{2}
\label{mu}
\end{equation}
with $f(\varepsilon)=1/(e^{\beta(\varepsilon-\mu)}+1)$.  
Here, we assume the Debye cutoff of 
$\omega_{\rm D}=2300$ K corresponding 
to the bond-stretching mode; i.e., the energy 
integral of eq.~(\ref{bcspi}) is from 
$\varepsilon=\mu-\omega_{\rm D}$ to 
$\mu+\omega_{\rm D}$.  
The calculated results for $T_c$ are shown in 
Fig.~\ref{cpatc1} as a function of $c_{\rm B}$ for 
various values of the attractive interaction $V$.  
We find that the value of $T_c$ increases rapidly 
with doping rate $c_{\rm B}$ and the attractive 
interaction $V$.  
If the energy of the bond-stretching mode of 
diamond $\omega=164$ meV is used as the cutoff 
energy in the integral of eq.~(\ref{bcspi}), 
the value of $T_c$ is found to decrease by $\sim$30\% 
but the general trend seen in Fig.~\ref{cpatc1} 
does not change.  

Our results may be compared with results of the 
first-principles LDA calculations based on either 
virtual crystal approximation \cite{boeri,lee} 
or supercell method \cite{lee,blase}.  
We find that our value of the density of states at 
the Fermi energy $\rho(\varepsilon_{\rm F})$ is 
about a factor 5 smaller than the values of the 
LDA calculations \cite{boeri} if we compare at 
doping rate of $c_{\rm B}=0.03$.  
The causes of this difference are the following; 
the absence of the three-fold degeneracy on one 
hand and the effects of disorder on the other 
hand in the present calculations.  
Actually, for the latter, we have the value 
$\rho(\varepsilon_{\rm F})=0.027$ $(0.046)$ at, e.g., 
$c_{\rm B}=0.01$ $(0.03)$ in our CPA calculation, 
while without disorder in the same model, we have 
$\rho(\varepsilon_{\rm F})=0.035$ $(0.053)$ at 
$c_{\rm B}=0.01$ $(0.03)$.  

The strength of the electron-phonon coupling constant 
$\lambda$ has also been estimated by the LDA 
calculations \cite{boeri,lee,blase}, where it was found 
that the contribution stems predominantly from the 
bond-stretching phonon mode and that the values are 
$\lambda\simeq 0.3-0.55$ at $c_{\rm B}\simeq 0.02-0.03$.  
The value of $T_c$ has thereby been estimated based 
on the BCS or McMillan's formula \cite{carbotte} and 
found to be in quantitative agreement with experiment 
\cite{boeri,lee,blase}.  
Then, from the comparison with the results of the LDA 
calculations, the strength of the attractive interaction 
$V$ in our theory may be estimated; i.e., by using 
the relation \cite{carbotte} 
\begin{equation}
\rho(\varepsilon_{\rm F})V=\frac{\lambda-\mu^*}{1+\lambda}
\end{equation}
with our calculated value of 
$\rho(\varepsilon_{\rm F})=0.027$ at $c_{\rm B}=0.01$ 
and a standard value of the Coulomb pseudopotential 
parameter $\mu^*\simeq 0.1$, and by assuming a typical 
value of the electron-phonon coupling constant 
$\lambda\simeq 0.31$ at $c_{\rm B}=0.01$, 
we can estimate the value of the attractive interaction 
to be $V\simeq 6$.  

\section{Comparisons with Experiment}

\subsection{Density of States}

The calculated results for the density of states 
(see Fig.~\ref{figdos}) may be compared with results 
of a number of spectroscopy experiments such as the 
photoemission and absorption spectra 
\cite{nakamura1,nakamura2,yokoya,ishizaka2} 
and optical conductivity spectra \cite{wu}.  
Although the effects of the matrix elements 
relevant to each experiment is not taken into 
account in the present calculations, the comparison 
can offer important information on the impurity 
states of the hole-doped diamond.  

We first point out that our calculated results for 
the density of states (see Fig.~\ref{figdos}) indicate 
that the impurity states evolve continuously by doping 
from the low-doping limit, where it has been 
established \cite{ramdas,kim} that the triply-degenerate 
impurity level exists for a single B impurity in 
otherwise pure diamond.  
This is in contrast to the claim of the rigid-band 
shift of the Fermi energy suggested in the 
first-principles LDA band calculations by the 
supercell method \cite{lee,blase}, which cannot 
reproduce the separated impurity level at low doping 
limit of diamond \cite{oguchi}.  

The presence or absence of the impurity states in 
the metallic regime of B-doped diamond should 
in principle be clarified by experiment.  
First, the angle-integrated photoemission spectroscopy 
experiment \cite{ishizaka2} has not so far provided 
evidences for the presence of the impurity 
states.  On the other hand, the X-ray 
absorption (XAS) and emission (XES) spectroscopy 
experiments \cite{nakamura1,nakamura2} 
seem to suggest the presence of the impurity states 
\cite{baskaran2} although not yet fully conclusive.  
The optical conductivity experiment \cite{wu} also 
seems to support the presence of the impurity states 
even in the metallic regime, where it has been found 
that, by doping, there appear the broad spectral 
features at $\sim 0.37$ eV that may correspond to 
the impurity states while the Drude weight is 
hardly visible.  

Further experimental studies are desired to clarify 
the nature of the impurity states in B-doped diamond.  

\subsection{Life Time and Mean-Free-Path}

The calculated results for the life time and 
mean-free-path are shown in Fig.~\ref{lifetime}.  
At $c_{\rm B}=0.01$, e.g., we find the values 
$E_{\rm F}\tau=0.48$ and $k_{\rm F}\ell=0.55$.  
These values of $E_{\rm F}\tau$ and $k_{\rm F}\ell$ 
are close to the experimental values 
$E_{\rm F}\tau=0.27$ 
and $k_{\rm F}\ell=0.32$, $0.48$, and $1.0$ for 
the three bands obtained from the line shape of 
quasiparticles by ARPES experiment for the highest 
$T_c$ sample \cite{yokoya}.  
The analyses of the measured resistivity 
\cite{takano1,bustarret,sidorov,ishizaka1} lead to 
similar small values of $E_{\rm F}\tau$ 
and $k_{\rm F}\ell$ as well.  

\begin{figure}[htb]
\begin{center}
\includegraphics[scale=0.83]{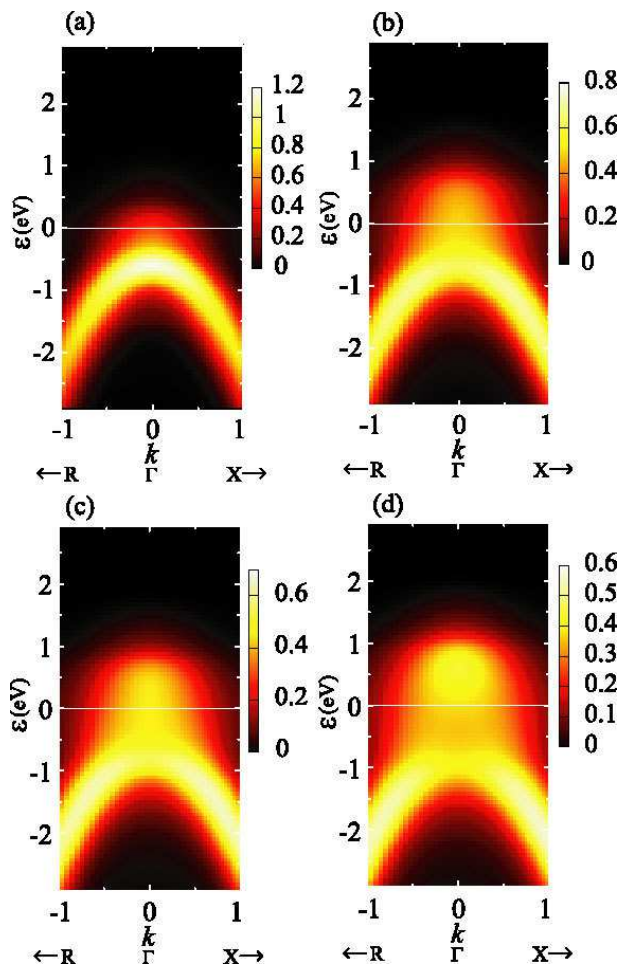}
\end{center}
\caption{(Color online) Calculated results for the 
single-particle spectra $\rho({\mib k},\varepsilon)$ 
at doping rates of (a) $c_{\rm B}=0.2\%$, 
(b) $c_{\rm B}=0.8\%$, (c) $c_{\rm B}=1.0\%$, 
and (d) $c_{\rm B}=1.5\%$.  
The finite resolutions in both energy (250meV) 
and momentum ($\pm$0.026au$^{-1}$) 
in experiment \cite{yokoya} are taken into account 
in the calculation.  The spectra below the Fermi 
energy (shown by horizontal line) may be 
compared with the experimental ARPES data obtained 
by Yokoya {\it et al.} \cite{yokoya}  The unit of 
energy is given in eV for comparison with 
experiment.  Horizontal axes are in units of $1/a$.}
\label{figarpes}
\end{figure}

\subsection{Angle-Resolved Photoemission Spectra}

The calculated results for the single-particle spectra 
in Fig.~\ref{figdispersion} may be compared with 
the experimental ARPES spectra given in Fig.~2 of 
ref.~\cite{yokoya}.  To do this, we recalculate the 
spectra by convoluting calculated results with 
the experimental resolutions in both energy 
(250 meV) and momentum ($\pm$0,026 au$^{-1}$) 
in view of the actual experiment \cite{yokoya}.  
The results near the Fermi energy are shown in 
Fig.~\ref{figarpes}, where the spectra both below 
(which can be compared with experiment) and above 
the Fermi energy are presented.  
We should stress again that the three-fold degeneracy 
of the top of the valence band is ignored in our 
calculations.  
We find however that the agreement between theory and 
experiment (below the Fermi energy) is obviously 
reasonable; 
i.e., even in the presence of the impurity states, 
the spectra of the valence band indeed look just as if 
the rigid-band shift of the Fermi energy were operating.  
But this apparent agreement does not imply that the 
impurity states are not present.  
This is in particular the case at low-doping regions; 
at $c_{\rm B}=0.2\%$, the impurity states are not clearly 
identified in the calculated angle-resolved spectra 
(see Fig.~\ref{figarpes}(a)), but in the calculated 
density of states (see Fig.~\ref{figdos}(a)), we find 
that the impurity states certainly exist, which are 
separated from the top of the valence band.  
Thus, up to the doping rates of $c_{\rm B}\simeq 1\%$, 
the shape of the angle-resolved spectra below the Fermi 
energy is not largely modified by the presence of the 
impurity states.  
We therefore consider that the claim \cite{yokoya} 
of the absence of the impurity states made from their 
ARPES data may not be justified, at least with 
resolutions in both momentum and energy used in 
the actual experiments.  Further experiments with 
much higher resolutions would be able to resolve 
the issue.  

\begin{figure}[hbt]
\begin{center}
\includegraphics[scale=0.37]{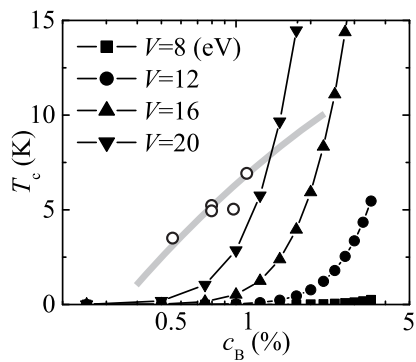}
\end{center}
\caption{Comparison between the calculated critical 
temperature $T_c$ as a function of $c_{\rm B}$ and 
the experimental data (open circles and broad curve) 
estimated by Mukuda {\it et al.} \cite{mukuda} from 
the analysis of the NMR spectra.  $T_c$ is given 
in units of K for comparison with experiment.  
The values of $V$ used for calculations are given 
in units of eV here.}
\label{cpatc2}
\end{figure}

\subsection{Superconducting Critical Temperature}

The calculated results for the doping dependence of 
$T_c$ are compared with experiment in Fig.~\ref{cpatc2}.  
We find that if we assume an appropriate value of $V$, 
e.g., $V\simeq 6$ (which corresponds to $\sim$20 eV), 
our result of $T_c\simeq 7$ K at $c_{\rm B}=1\%$ is 
in reasonable agreement with results of the LDA 
calculations \cite{boeri,lee,blase} and therefore 
with experiment.  
In particular, we find that the rapid increase 
in $T_c$ with increasing doping rate, which is 
observed in experiment, is reproduced fairly well.  

If seen more quantitatively, however, one may notice 
that the calculated concave dependence of 
$T_c$ on $c_{\rm B}$ (see Fig.~\ref{cpatc2}) is 
somehow not consistent with experiments where the 
nearly linear dependence on $c_{\rm B}$ in the 
logarithmic scale is seen.  We consider that the 
discrepancy may partly be due to the insufficiency 
(or mean-field nature) of CPA.  
In CPA, both the width and height of the density 
of states of the impurity states vary as 
$\sim\sqrt{c_{\rm B}}$ (hence, the area varies 
as $\sim c_{\rm B}$), whereas in reality, there 
is a tendency \cite{taylor,alben1,alben2} that the 
width varies as $\sim c_{\rm B}$ and the height 
is almost kept constant $\sim 1$.  
If we take into account this tendency in our 
calculations, the agreement between theory and 
experiment will be improved.  

\section{Discussions}

In this section, we want to discuss some issues 
and possibilities.  

First, let us consider possible improvements of our 
calculations presented in this paper.  We should 
first of all point out that our calculated values 
of $T_c$ are not of sufficient quantitative 
significance since the calculations are based on 
the single-band model of the simple-cubic lattice 
and the instantaneous interaction for the attractive 
force.  For more quantitative estimation of $T_c$, 
we may need the following improvements: 
(i) The nearly-degenerate three-band structure 
around the top of the valence band of diamond 
in the actual lattice structure should be included.  
(ii) The retardation effects for the electron-phonon 
coupling should be taken into account.  
(iii) The dynamical aspect of the Coulomb interaction 
between holes should be important; due to the effects 
of diffusion by randomness, the Coulomb pseudopotential 
parameter $\mu^*$ should be effectively larger 
\cite{maekawa,graybeal}.  
(iv) Approximations beyond CPA will be desirable 
as pointed out in \S 4.4.  
These subjects deserve further studies.  

\begin{figure}[htb]
\begin{center}
\includegraphics[scale=0.47]{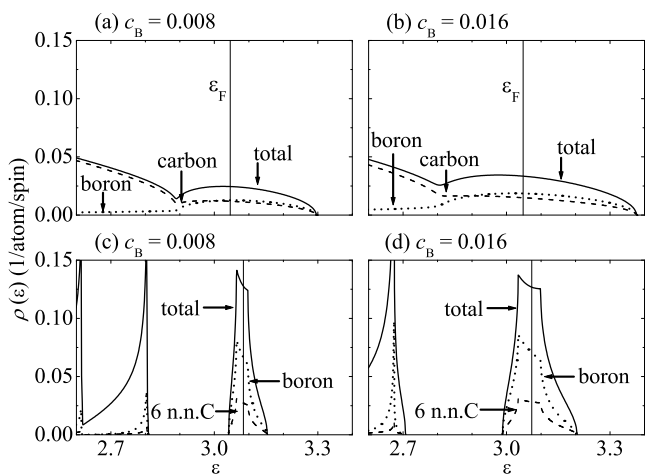}
\end{center}
\caption{Densities of states near the Fermi energy 
obtained by CPA (upper panels) and those calculated 
under the assumption of the periodic arrangement 
of B ions (lower panels).  The $5\times 5\times 5$ 
and $4\times 4\times 4$ supercells are assumed in 
(c) and (d), respectively.  
Partial densities of states are also shown 
where `6 n.n.C' means the partial density of states 
summed over the six nearest-neighbor carbon sites 
around a boron ion.  Vertical lines indicate the 
Fermi energy $\varepsilon_{\rm F}$.}
\label{supercelldos}
\end{figure}

Next, let us consider what happens to $T_c$ if 
there were no effects of disorder of B ions.  
To see this, we carry out the supercell calculations 
for the same set of parameter values used in the 
CPA calculations.  The B ions are arranged 
periodically by using  
$6\times 6\times 6$, 
$5\times 5\times 5$, 
$4\times 4\times 4$, and 
$3\times 3\times 3$ 
supercells of the simple-cubic lattice, 
which correspond to 
0.46\%, 
0.80\%,
1.56\%, and 
3.7\% doping rates, respectively.  
The calculated results for the density of states 
are shown in Fig.~\ref{supercelldos}, which are 
compared with the results of CPA at the same 
doping rates.  
We find that a rather narrow density of states 
for the impurity states appears above the top of 
the valence band, which form bands (impurity band), 
and that the density of states at the Fermi energy 
is much higher than the result of the CPA 
calculation.  The separation between the impurity 
band and the top of the valence band is $\sim 1$ eV, 
which is significantly larger than the energy of 
the impurity level in the dilute limit, $0.37$ eV, 
due to level repulsions.  

\begin{figure}[htb]
\begin{center}
\includegraphics[scale=0.38]{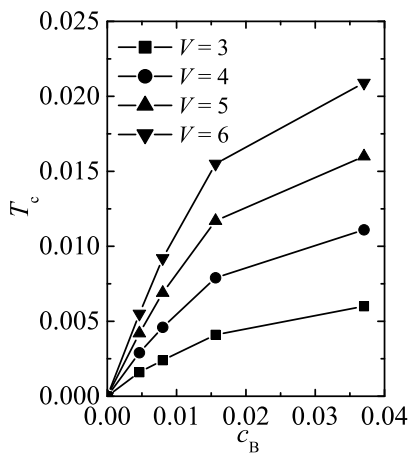}
\end{center}
\caption{Superconducting critical temperature $T_c$ 
obtained under the assumption of the periodic 
arrangement of B ions.}
\label{supercelltc}
\end{figure}

We then calculate the values of $T_c$ for these 
supercell systems by using eq.~(\ref{supertc}).  
The results are shown in Fig.~\ref{supercelltc}.  
We find that the obtained values of $T_c$ are nearly 
an order of magnitude higher than the results of 
the CPA calculations for the same attractive 
strength (see Fig.~\ref{cpatc1}).  
Thus, we may conclude that, if impurity B ions 
could be arranged periodically, one would be able 
to have the critical temperature of an order of 
$\sim$100 K.  This value of $T_c$ exceeds 
$T_c\simeq 30-40$ K of what is called the theoretical 
limit of the BCS theory for conventional 
superconductors; this is not surprising because the 
Debye frequency of diamond $\omega_{\rm D}=2300$ K 
is exceptionally large, together with the presence 
of the strong electron-phonon coupling because of 
the hole doping into the covalent bonds \cite{fukuyama0}.  
In other words, we can say that, in actual B-doped 
diamond, the effect of disorder indeed strongly 
suppresses the value of $T_{\rm c}$.  
Note the `convex' dependence of $T_c$ on $c_{\rm B}$ 
in Fig.~\ref{supercelltc}, which is in contrast to 
that of the CPA calculation where the dependence is 
`concave' (see Fig.~\ref{cpatc1}); i.e., the $T_c$ 
in the supercell calculations decreases very rapidly 
with decreasing $c_{\rm B}$.  
This comes from the narrow density of states of 
the impurity band, where the width of the impurity 
band becomes less than the Debye cutoff and 
the integral in eq.~(\ref{supertc}) becomes very small.  

Together with reduction of the effects of disorder 
to raise $T_c$, the increase of hole concentration 
is obviously another way.  
In the actual experiment, one may try to find 
out way to remove or reduce the H atoms 
introduced in the synthetic processes of the 
materials; i.e., the reduction of passivation rate.  

Finally, we want to point out the possibility of the 
occurrence of superconductivity in similar 
covalent-bonded materials.  Based on the present 
study, we may suggest that the doping of carriers 
in boron-nitride (BN) and silicon-carbide (SiC) 
may be quite interesting; these materials with 
the cubic (zinc-blende) structure not only have 
the valence-band structure with a large band gap 
similar to diamond but also have the bond-stretching 
phonon mode with very high frequency 
\cite{sanjurjo,gubanov,rodriguez,karch,lebedev,
kuwabara,persson}.  
Then, we may expect that the doping of enough 
carriers would lead to superconductivity with 
appreciable $T_c$.  

\section{Conclusions}

We have studied the superconductivity in B-doped 
diamond by use of CPA for treating the substitutional 
disorder of B ions and by treating the attractive force 
between holes by the ladder approximation under the 
assumption of instantaneous interaction.  
The main messages obtained from our studies are the 
following: 

(i) The effects of disorder are essential for 
understanding the electronic state of B-doped diamond.  
The broadening of the density of states and single-particle 
spectra near the Fermi energy, as well as the very short 
life-time and mean-free-path, are due to the effects of 
disorder, i.e., the large imaginary part of the self-energy 
induced by the disorder of B ions.  

(ii) The pairing susceptibility derived by CPA takes 
the form of the BCS theory, where the effects of disorder 
are renormalized solely into the density of states, even 
in the presence of strong disorder that deforms the shape 
of the density of states.  The vertex correction is 
found to be essential here: i.e., the susceptibility 
becomes too small if we take into account only the effect 
of the self-energy, but it is recovered by the vertex 
correction, leading to the BCS expression for the 
pairing susceptibility.  

(iii) The superconducting critical temperature $T_c$ 
can be raised substantially by reducing the effects of 
disorder in doping processes, e.g., by controled 
spatial arrangement of B ions.  

We believe that the doped semiconductors should have 
capabilities for producing a new class of superconducting 
materials and therefore we hope that our work presented 
here will encourage further experimental studies of 
superconductivity induced by doping into covalent-bonded 
semiconductors.  

\section*{Acknowledgment}
One of the authors (H.F.) thanks G. Baskaran for initial 
input and various discussions and M. Dresselhaus for 
informative discussion regarding C-related materials.  
Discussions on experimental facts with S. Shin in 
early stage are also acknowledged.  
Thanks are also due to E. Bustarret, K. Ishizaka, 
K. Miyake, J. Mizuki, H. Mukuda, J. Nakamura, 
T. Oguchi, Y. Takada, Y. Takano, N. Yamada, 
and T. Yokoya.  
T.~S. acknowledges support from Mochizuki Foundation.  
This work was supported in part by Grants-in-Aid for 
Scientific Research (Nos.~16076201 and 18540338) 
from the Ministry of Education, Science, Sports, and 
Culture of Japan. 
A part of computations was carried out at the Research 
Center for Computational Science, Okazaki Research 
Facilities, and the Institute for Solid State Physics, 
University of Tokyo.


\begin{thebibliography}{99} 


\bibitem{ekimov} E. A. Ekimov, V. A. Sidorov, E. D. Bauer, 
N. N. Mel'nik, N. J. Curro, J. D. Thompson, 
and S. M. Stishov: Nature (London) \textbf{428} (2004) 542.  

\bibitem{takano1} Y. Takano, M. Nagao, I. Sakaguchi, 
M. Tachiki, T. Hatano, K. Kobayashi, H. Umezawa, 
and H. Kawarada: Appl. Phys. Lett. \textbf{85} (2004) 2851.  

\bibitem{hoesch} M. Hoesch, T. Fukuda, T. Takenouchi, 
J. P. Sutter, S. Tsutsui, A. Q. R. Baron, M. Nagao, 
Y. Takano, H. Kawarada, and J. Mizuki: cond-mat/0512424.  

\bibitem{takano2} Y. Takano, M. Nagao, T. Takenouchi, 
H. Umezawa, I. Sakaguchi, M. Tachiki, and H. Kawarada: 
Diam. Relat. Matter. \textbf{14} (2005) 1936.  

\bibitem{umezawa} H. Umezawa, T. Takenouchi, Y. Takano, 
K. Kobayashi, M. Nagao, I. Sakaguchi, M. Tachiki, 
T. Hatano, G. Zhong, and M. Tachiki: cond-mat/0503303.  

\bibitem{mukuda} H. Mukuda, T. Tsuchida, A. Harada, 
Y. Kitaoka, T. Takenouchi, Y. Takano, M. Nagao, 
I. Sakaguchi, and H. Kawarada: to be published in 
{\it Science and Technology of Advanced Materials} 
(Elsevier, 2006).  

\bibitem{baskaran1} G. Baskaran: cond-mat/0404286, 
to appear in J. Supercond. (2006).  

\bibitem{baskaran2} G. Baskaran: cond-mat/0410296. 

\bibitem{bustarret} E. Bustarret, J. Ka\v cmar\v cik, C. Marcenat, 
E. Gheeraert, C. Cytermann, J. Marcus, and T. Klein: 
Phys. Rev. Lett. \textbf{93} (2004) 237005.  

\bibitem{sidorov} V. A. Sidorov, E. A. Ekimov, 
S. M. Stishov, E. D. Bauer, and J. D. Thompson: 
Phys. Rev. B \textbf{71} (2005) 060502(R).  

\bibitem{ishizaka1} K. Ishizaka, R. Eguchi, S. Tsuda, 
T. Yokoya, T. Kiss, T. Shimojima, T. Togashi, 
S. Watanabe, C.-T. Chen, C. Q. Zhang, Y. Takano, 
M. Nagao, I. Sakaguchi, T. Takenouchi, H. Kawarada, 
and S. Shin: cond-mat/0604370.  

\bibitem{yokoya} T. Yokoya, T. Nakamura, T. Matsushita, 
T. Muro, Y. Takano, M. Nagao, T. Takenouchi, 
H. Kawarada, and T. Oguchi: 
Nature (London) \textbf{438} (2005) 648.  

\bibitem{fukuyama0} H. Fukuyama: J. Supercond. Novel Magn., 
to appear (2006).  

\bibitem{ramdas} A. K. Ramdas and S. Rodriguez: 
Rep. Prog. Phys. \textbf{44} (1981) 1297.  

\bibitem{kim} H. Kim, A. K. Ramdas, S. Rodriguez, 
M. Grimsditch, and T. R. Anthony: 
Phys. Rev. Lett. \textbf{83} (1999) 4140.  

\bibitem{boeri} L. Boeri, J. Kortus, and O. K. Anderson: 
Phys. Rev. Lett. \textbf{93} (2004) 237002.  

\bibitem{lee} K. W. Lee and W. E. Pickett: 
Phys. Rev. Lett. \textbf{93} (2004) 237003; 
Phys. Rev. B \textbf{73} (2006) 075105.  

\bibitem{blase} X. Blase, C. Adessi, and D. Conn\' etable: 
Phys. Rev. Lett. \textbf{93} (2004) 237004.  

\bibitem{nakamura1} J. Nakamura, E. Kabasawa, N. Yamada, 
Y. Einaga, D. Saito, H. Issiki, S. Yugo, and 
R. C. C. Perera: Phys. Rev. B \textbf{70} (2004) 245111. 

\bibitem{nakamura2}  J. Nakamura, T. Oguchi, N. Yamada, 
K. Kuroki, K. Okada, Y. Takano, M. Nagao, I. Sakaguchi, 
H. Kawarada, R. C. C. Perera, and D. L. Ederer: 
cond-mat/0410144.  

\bibitem{wu} D. Wu, Y. C. Ma, Z. L. Wang, Q. Luo, C. Z. Gu, 
N. L. Wang, C. Y. Li, X. Y. Lu, and Z. S. Jin: 
Phys. Rev. B \textbf{73} (2006) 012501.  

\bibitem{oguchi} T. Oguchi: private communication.  

\bibitem{xiang} H. J. Xiang, Z. Li, J. Yang, J. G. Hou, 
and Q. Zhu: Phys. Rev. B \textbf{70} (2004) 212504.  

\bibitem{ohta} Y. Ohta, T. Shirakawa, S. Horiuchi, 
and H. Fukuyama: to appear in Proceedings of M$^2$S-HTSC, 
Dresden, 2006.  

\bibitem{VKE} B. Velicky, S. Kirkpatrick, and H. Ehrenreich:  
Phys. Rev. \textbf{175} (1968) 747.  

\bibitem{yonezawa} F. Yonezawa: Prog. Theor. Phys. 
\textbf{40} (1968) 734. 

\bibitem{elliott} R. J. Elliott, J. A. Krumhansl, and P. L. Leath: 
Rev. Mod. Phys. \textbf{46} (1974) 465.  

\bibitem{fukuyama1} H. Fukuyama: Phys. Rev. B \textbf{5} (1972) 2872. 

\bibitem{fukuyama2} H. Fukuyama: Phys. Rev. B \textbf{8} (1973) 4288.  

\bibitem{anderson} P. W. Anderson: J. Phys. Chem. Solids 
\textbf{11} (1959) 26.  

\bibitem{ishizaka2} K. Ishizaka, R. Eguchi, S. Tsuda, 
T. Kiss, T. Shimojima, T. Yokoya, S. Shin, 
T. Togashi, S. Watanabe, C.-T. Chen, C. Q. Zhang, 
Y. Takano, M. Nagao, I. Sakaguchi, T. Takenouchi, 
and H. Kawarada: to be published in 
{\it Science and Technology of Advanced Materials} 
(Elsevier, 2006).  

\bibitem{carbotte} J. P. Carbotte: 
Rev. Mod. Phys. \textbf{62} (1990) 1027.  

\bibitem{taylor} D. W. Taylor: Phys. Rev. \textbf{156} 
(1967) 1017.  

\bibitem{alben1} R. Alben, M. Blume, H. Krakauer, 
and L. Schwartz: Phys. Rev. B \textbf{12} (1975) 4090.  

\bibitem{alben2} R. Alben, M. Blume, and M. McKeown: 
Phys. Rev. B \textbf{16} (1977) 3829.  

\bibitem{maekawa} S. Maekawa and H. Fukuyama: 
J. Phys. Soc. Jpn. \textbf{51} (1981) 1380.  

\bibitem{graybeal} J. M. Graybeal and M. R. Beasley: 
Phys. Rev. B \textbf{29} (1984) 4167.  

\bibitem{sanjurjo} J. A. Sanjurjo, E. Lopez-Cruz, 
P. Vogi, and M. Cardona: 
Phys. Rev. B \textbf{28} (1983) 4579. 

\bibitem{gubanov} V. A. Gubanov, E. A. Pentaleri, 
C. Y. Fong, B. M. Klein: Phys. Rev. B \textbf{56} 
(1997) 13077. 

\bibitem{rodriguez} P. Rodriguez-Hernandez, 
M. Gonzales-Diaz, A. Munoz: 
Phys. Rev. B \textbf{51} (1995) 14705. 

\bibitem{karch} K. Karch, P. Pavone, W. Windl, 
O. Schutt, and D. Strauch: Phys. Rev. B \textbf{50} 
(1994) 17054. 

\bibitem{lebedev} A. A. Lebedev: Semiconductors 
\textbf{33} (1999) 107. 

\bibitem{kuwabara} H. Kuwabara and S. Yamada: 
Phys. Stat. Solidi A \textbf{30} (1975) 739. 

\bibitem{persson} C. Persson and U. Lindefelt: 
J. Appl. Phys. \textbf{82} (1997) 5496. 

\end{thebibliography}
\end{document}